# Optical Readout Studies of the Thick-COBRA Gaseous Detector

F. Garcia,[a,*] F. M. Brunbauer,[b] M. Lisowska,[b,c] H. Müller,[b] E. Oliveri,[b]
D. Pfeiffer,[b,d] L. Ropelewski,[b] J. Samarati,[b,d] F. Sauli,[b] L. Scharenberg,[b,e]
A. L. M. Silva,[f] M. van Stenis,[b] R. Veenhof,[g] J. F. C. A. Veloso[f]

[a] *Helsinki Institute of Physics, University of Helsinki,*
    *00014 University of Helsinki, Finland*

[b] *CERN,*
    *385 Route de Meyrin, 1217 Meyrin, Geneva, Switzerland*

[c] *Wrocław University of Science and Technology,*
    *Wybrzeże Wyspiańskiego 27, 50-370 Wrocław, Poland*

[d] *European Spallation Source (ESS ERIC),*
    *P.O. Box 176, SE-22100 Lund, Sweden*

[e] *University of Bonn,*
    *Regina-Pacis-Weg 3, 53113 Bonn, Germany*

[f] *I3N-Physics Department of University of Aveiro,*
    *3810-193 Aveiro, Portugal*

[g] *National Research Nuclear University MEPhI (Moscow Engineering Physics Institute),*
    *Kashirskoe highway 31, Moscow, 115409, Russia*

    E-mail: `Francisco.Garcia@helsinki.fi`

ABSTRACT: The performance of a Thick-COBRA (THCOBRA) gaseous detector is studied using an optical readout technique. The operation principle of this device is described, highlighting its operation in a gas mixture of Ar/CF$_4$ (80/20%) for visible scintillation light emission. The contributions to the total gain from the holes and the anode strips as a function of the applied bias voltage were visualized. The preservation of spatial information from the initial ionizations was demonstrated by analyzing the light emission from 5.9keV X-rays of an $^{55}$Fe source. The observed non-uniformity of the scintillation light from the holes supports the claim of a space localization accuracy better than the pitch of the holes. The acquired images were used to identify weak points and sources of instabilities in view of the development of new optimized structures.

[*] corresponding author

# Contents

## 1. Introduction

The Thick-COBRA gaseous detector, hereafter THCOBRA [1], was studied with optical readout to examine and visualize its operation. By recording secondary scintillation light, the location of the avalanche multiplication could be visualized with high granularity permitting to compare the response of the detector in light-collection mode with what is observed in charge-collection mode.

The THCOBRA detector is based on the Gas Electron Multiplier (GEM) [2] technology and can be described as a hybrid amplification structure combining a Thick-GEM (THGEM) [3][4] and a Micro-Strip Gaseous Counter (MSGC) [5]. As such, it is an evolution of the Micro-Hole Strip Plate (MHSP) [6] using a more robust substrate. It has electrodes on the top and bottom sides of a thick substrate. The bottom electrode is segmented to feature thin anode strips in between the GEM holes.

In recent years, the optical readout of scintillation light emitted during avalanche multiplication has been widely used as a readout modality for Micro-Patterns Gaseous Detectors (MPGDs). Stable operation at large gas gains in GEMs [2] and Micromegas [7], as well as the discovery of gas mixtures with high scintillation light yield in the visible wavelength region have made optical readout more accessible with the use of recent developments of high performance digital imaging sensors. Advances in commercially available cameras with high-resolution and low-noise sensors have allowed for unprecedented image quality and sensitivity, making optical readout an attractive readout modality for gaseous radiation detectors.

Combining state-of-the-art imaging sensors with MPGDs operated in well-understood scintillating gas mixtures [8] has become a powerful tool for studying the performance and operation principles of different types of MPGDs. Stable operation of multi-stage amplification structures at large gas gains has been successfully implemented for particle tracking with GEM [9] detectors and in GEM-based TPCs [10][11][12] achieving good image quality in terms of high signal to noise ratio for tracks crossing their sensitive volumes.

Applications such as X-ray imaging triple-GEM [13] and Micromegas detectors have also shown high spatial resolution [14]. In the present study, the operation concept of the THCOBRA detector was visualized by optical readout:

- Light and charge measurements, in order to visualize sharing between the THGEM and the MSGC-like structures of this hybrid device.

- Corroborate in light-collection mode, the preservation of the primary ionization position information already observed in charge-collection mode.

- Demonstrate the detection of low-energy X-rays from an $^{55}$Fe radioactive source.

- Localize weak spots of the THCOBRA structure causing instabilities at large gains, in order to improve its design.

The principles of operation are described highlighting the two amplification stages. The experimental setup is shown along with the integration of the THCOBRA foil into a gaseous detector with an optical readout. Finally, the measurements results are discussed together with conclusions.

## 2. Description of THCOBRA and Principle of Operation

In the Fig. 1, a picture of the THCOBRA foil with its electrodes' pattern on the bottom and top sides [15] is shown.

The foil with an area of 10x10cm$^2$ was manufactured on a printed circuit G10 board with a thickness of 400μm and was cladded with 50μm thick copper strips on both sides. The top electrode has strips of 400μm width, while the bottom side has cathode strips of 200μm width around the holes and anode strips of varying widths between hole. The holes diameter is 300μm with a pitch of 1mm. The principle of operation of THCOBRA foil is based on two amplification stages (Fig. 2) [16]. This hybrid gaseous electron multiplier works as a THGEM inside the holes and as a MSGC-like structure between cathode and anode strips to achieve higher gas gain and consequently stronger light emission.

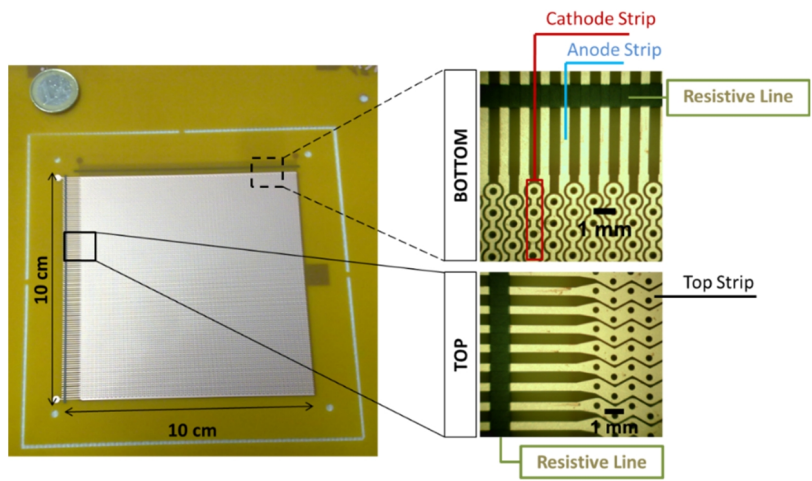

**Figure 1.** THCOBRA layout. View of the foil on the left. Bottom and top electrodes on the right. The upper-right picture displays cathode strips around the holes and anode strips inbetween the holes. The top electrode (bottom-right) has segmented strips hereafter called top contacts.

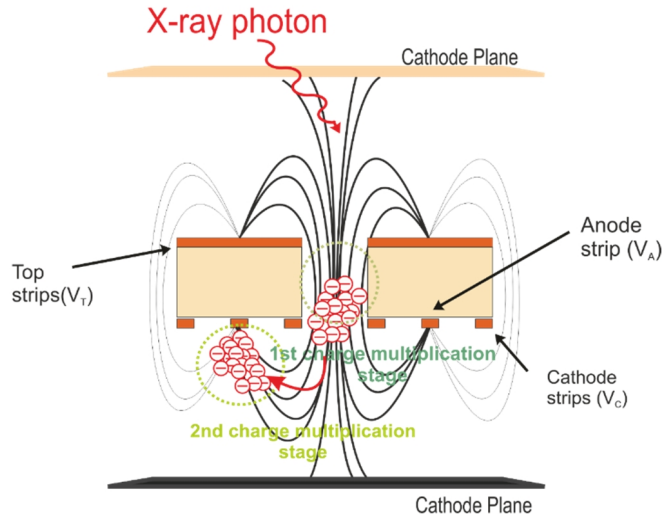

**Figure 2.** The operation principle of THCOBRA foil is based on a two-stage amplification structure, where the first amplification occurs within the holes and the second one near the anode strips.

To operate the structure, a voltage difference between the top contact and the cathode is applied to create high electric field strength inside the holes as is the case for THGEMs. In addition, between the anode and cathode strips a voltage difference is applied as in a MSGC to create a high electric field at the anode strips and add its contribution to the total gas gain. In stable operation, such a structures can reach gas gains of the order of $10^4 - 10^5$ in different gas mixtures [17] as has been previously reported.

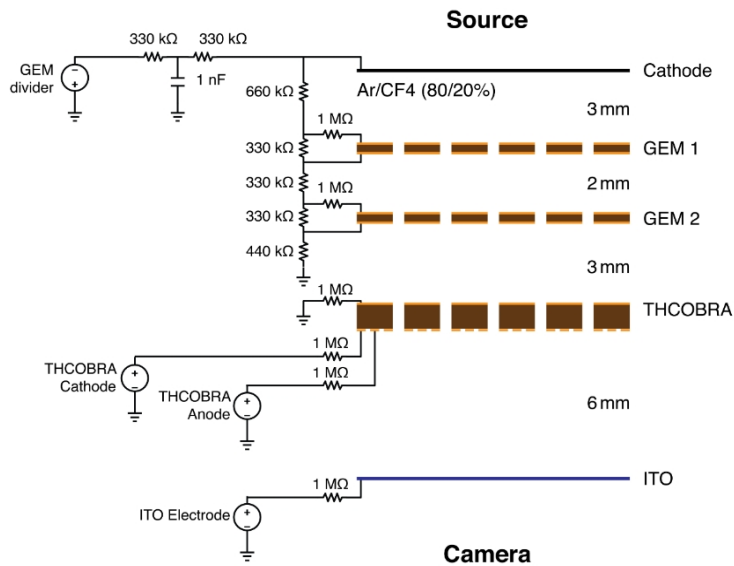

**Figure 3.** Schematics of the detector. A THCOBRA foil, a double-GEM stack, a cathode (on top) and an ITO electrode (on bottom) make up the optically read out detector.

## 3. Experimental Setup

The detector used in this study consists of a double GEM stack (two standard thin GEM foils) on top of the THCOBRA foil, used as a pre-amplification stage to achieve higher gas gain. On the source side, a copper cathode foil is used to create a drift field in the 3 mm thick conversion gap. A camera was placed outside of the gas volume facing the bottom of the THCOBRA foil through a glass plate coated with 25nm thick Indium-Tin Oxide (ITO). This ITO electrode is electrically conductive and has an optical transparency of approximately 80% in the visible wavelength region.

The detector was powered as shown in Fig. 3 using multiple independent high voltage lines, with one of them connected to a passive divider for biasing the double-GEM stack. The top electrode of the THCOBRA was grounded and two high voltage lines were used to bias the bottom electrodes: one for the cathode and another one for the anode. This allowed for independent control of the two amplifications stages of the hybrid structure.

An Ar/$CF_4$ gas mixture with volume concentrations of 80% and 20% respectively, was used for the present study. This allows to have the initial interaction dominated by Ar atoms, while the $CF_4$ provides visible scintillation light emission. The scintillation light emission spectrum for this mixture features wide emission bands at Ultraviolet (UV) and visible wavelengths, which are attributed to $CF_4$ [8]. The light emission spectrum matches well the quantum efficiency of conventional Charge-Coupled Device (CCD) and Complementary Metal Oxide Semiconductor (CMOS) imaging sensors. In addition, the broad emission lines of the scintillation spectrum also contain atomic emission lines in the near-infrared range attributed to Ar [18].

In Fig. 4 the experimental setup is shown with the THCOBRA detector in between an X-ray generator and the CCD camera. The distance between the camera and the detector is determined by the focal distance of the optical readout system. Supplementary lenses were added to reduce the minimum object distance. In the present setup, the minimum object distance is approximately 8 cm for a lens with 50mm focal length and additional collimating lenses. The cameras used for these tests were an Electron Multiplying Charge-Coupled Device (EM-CCD), which has high sensitivity and a 6-megapixel CCD camera for high resolution studies[*].

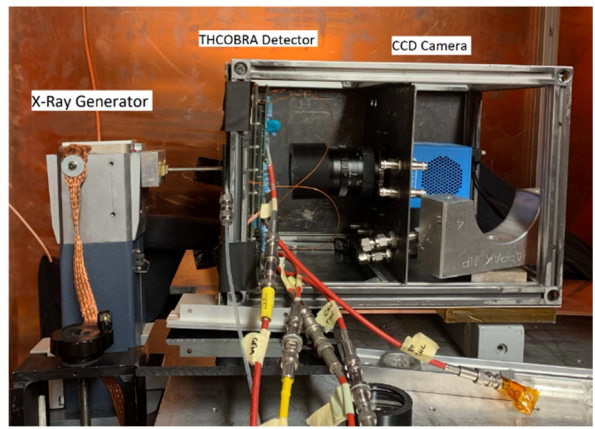

**Figure 4.** Experimental setup consisting of an X-ray generator, the THCOBRA gaseous detector and a high-resolution CCD camera.

---

[*] The high-resolution camera was the QImaging Retiga R6 [19] which featured a 12.5×10mm$^2$ imaging sensor with 6M pixels of 4.54×4.54μm$^2$. For high sensitivity measurements an ImagEM X2 [20] EMCCD camera was used featuring an image sensor of 8.19x8.19mm$^2$ with 512x512 pixels of 16x16μm$^2$.

## 4. Measurement Results

### 4.1 Contributions from amplification stages

The light emission from the two amplifications stages of the THCOBRA foil was visualized as a function of the applied bias voltages. The foil was biased by increasing the voltage difference across the holes up to the maximum achievable gain in stable operation conditions. This was done by keeping the cathode and anode located at the bottom of the THCOBRA foil at the same potential. Fig. 5 shows the powering scheme used for the measurement.

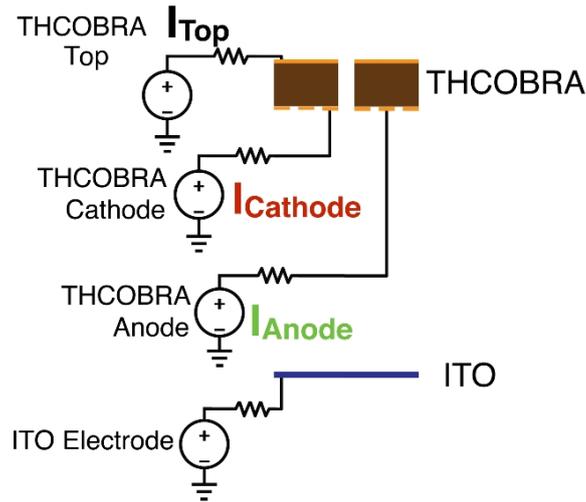

**Figure 5.** Schematics of the measurement setup for the visualization of the light sharing between the two amplification stages of the THCOBRA.

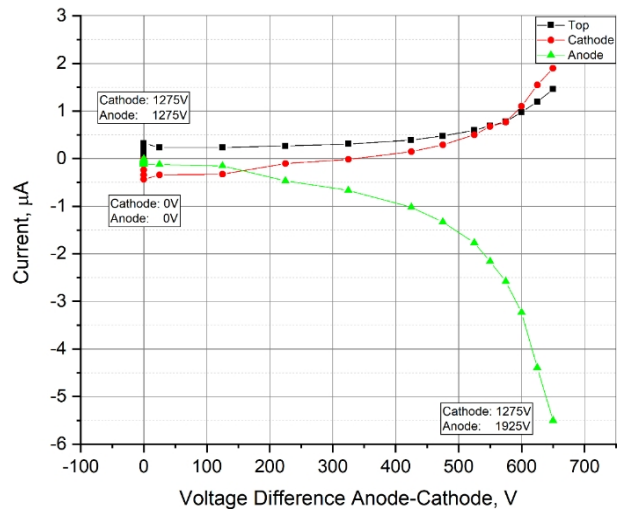

**Figure 6.** The currents of the THCOBRA electrodes as a function of the voltage difference between anode and cathode strips. The current on the top contact (square), the cathode (circle) and the anode (triangle) is shown.

The top contact and the cathode (in Fig. 5.) were connected to separated high voltage lines and the currents were monitored. Similarly, the anode strips were biased and monitored independently. The measured currents are presented in Fig. 6.

At zero voltage difference between anode strips and the cathode contact ($V_{AC} = 0V$), the multiplication occurs only in the holes and creates a small current on the cathode (from the arrival of electrons from the avalanche) and top strips i.e. only the holes contribute to the gas gain. By increasing this voltage difference, a negative polarity current at the anode starts to increase together the currents at the top contact and the cathode (from the ions produced in the avalanche). This overall effect indicates the onset of amplification from both stages and their contributions to the total gas gain. In addition to the current measurements explained above, images shown in Fig.7 were recorded the applied voltage difference. The $V_{AC}$ was increased from 0 V to 650V. The voltage across the holes was kept constant at 1200V, while biasing the cathode at 1200V with the top contact grounded. For low absolute values of $V_{AC}$, only the holes provide amplification and therefore light is emitted only from the holes. Increasing $V_{AC}$ leads to the onset of gas gain at the anode strips which begin to contribute to the total gas gain. This is visible by the light emitted from the anode strips. These observations are in agreement with the interpretation given by the current measurements shown above in Fig.6.

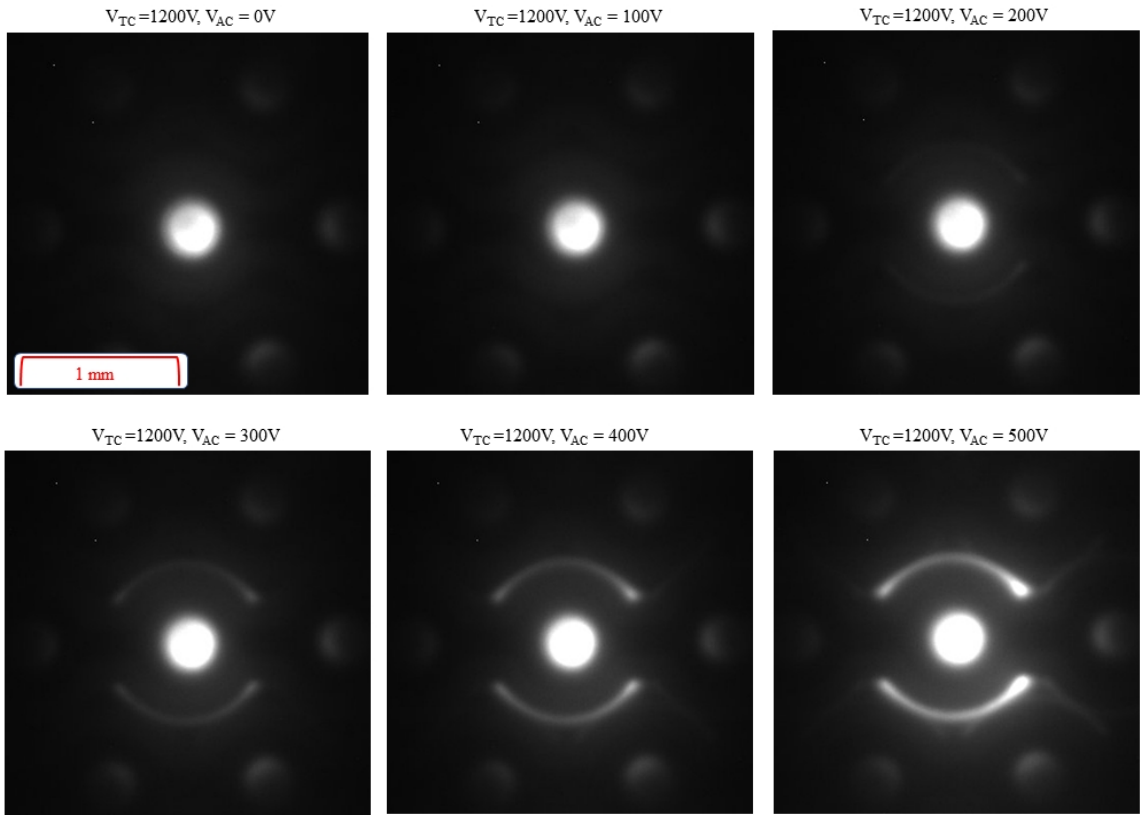

**Figure 7.** Recorded images of the THCOBRA light emission as a function the anode-cathode voltage difference. $V_{TC}$ is the voltage across the THGEM hole and $V_{AC}$ is the voltage difference between anode and cathode strips.

Light is emitted primarily from the edges and corners of the anode strips when $V_{AC}$ is sufficiently high for multiplication to occurs at the strips. This is attributed to an increased concentration of electric field lines at those locations.

## 4.2 Preservation of spatial information during amplification

Measurements in charge-collection mode have previously reported spatial resolution values below 0.5mm for THCOBRA detectors, which is well below the 1mm pitch of the holes [21]. This suggests a preservation of position information from primary ionizations during the avalanche multiplication i.e. a so-called "memory effect". Optical measurements were used to investigate whether the position of the avalanches can be localized within the holes.

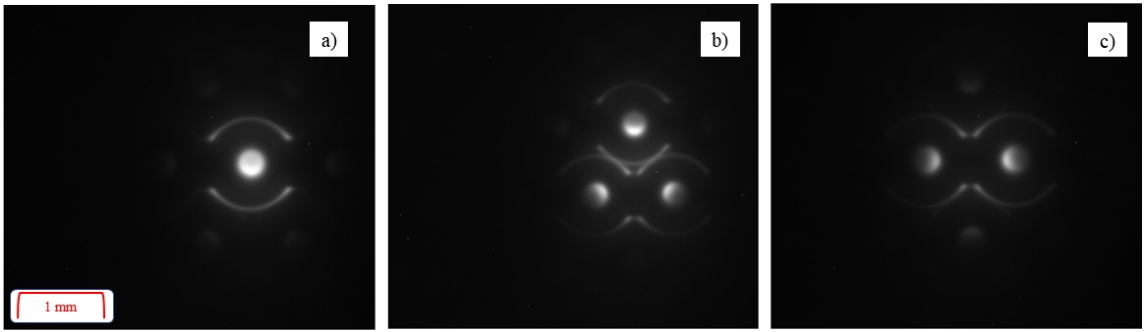

**Figure 8.** Illumination with a collimated X-rays beam of a single cell of the THCOBRA. The beam is a) centred in the middle of a hole and subsequently is b) centred in the middle of three holes and c) inbetween two holes.

The THCOBRA detector was irradiated with a collimated X-ray generator as shown in Fig. 8. The beam was placed in three different locations: in the middle of a single hole (Fig. 8a), in the middle of three holes (Fig. 8b) and in the middle between two holes (Fig. 8c). As shown in Fig. 8, there is an asymmetry in the light recorded from holes and anode strips indicating that spatial information might well be preserved in the avalanches. When the beam is centered on a single hole (Fig. 8a), only this hole and the anode strips in the vicinity of that hole contribute to the observed charge multiplication and emitted scintillation light. In the recorded image, light is emitted almost uniformly from the hole, with a slight shift towards the bottom which may be attributed to imperfect centering of the X-ray beam with respect to the hole. Both neighboring anode strips are contributing equally with their corners displaying increased light intensity compared to the central regions of the strips. When the X-ray beam was centred between three holes, all of them were contributing and emitting light (Fig. 8b). Within each hole, the sides facing towards the incident beam position (center between the three holes) were observed to emit significantly more light than the outer regions of those holes. The same was observed for the anode strips, where only segments of the anode strips in the vicinity of the center between the three holes were contributing strongly to the observed light emission and almost no light was observed from the outward anode strips. The same was observed for the case of the X-ray beam centered inbetween two holes (Fig. 8c), where the amplification and emitted scintillation light was shared between both of them. The observed light emission for these three cases confirms that spatial information of the incident electrons is retained during the avalanche multiplication of the THCOBRA foil. The precise implications of this effect on the achievable spatial resolution of this detector could not be determined with the presented setup, since the two thin GEMs used for pre-amplification above the THCOBRA foil resulted in a widened electron cloud on top of it. Thus, some of the width of the observed light emission profiles might well be a result from the multiplication in the two thin GEMs or from diffusion during the transfer processes. Nevertheless, the observation that only certain regions within the hole or along the anode strips contribute to the observed light emission shows that spatial information of electrons reaching the THCOBRA foil is not completely lost during the avalanche multiplication in this structure.

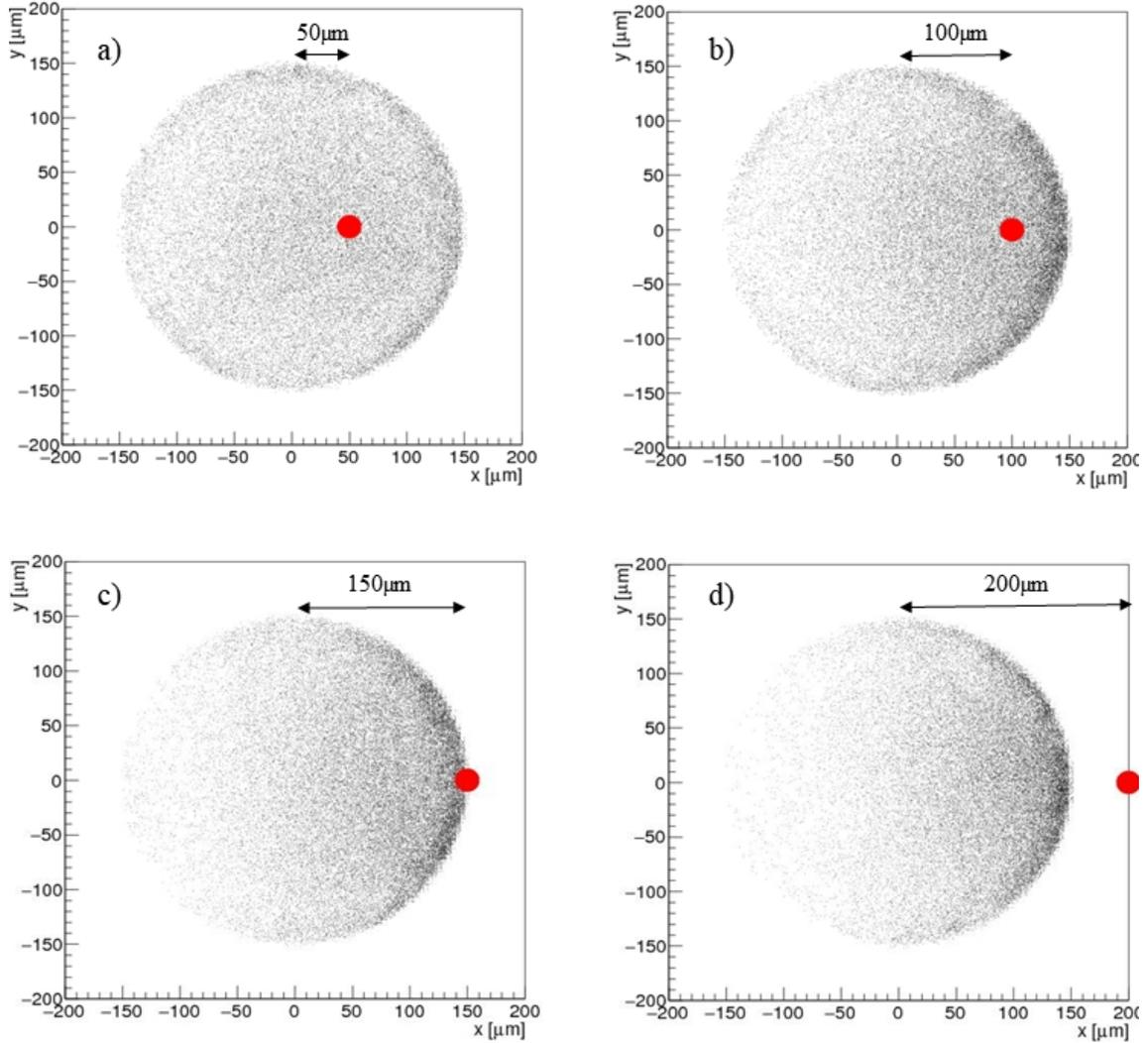

**Figure 9.** Garfield simulations of the avalanches created by primary electrons. The red point indicates the entry point of the electrons at different distances from the center of the hole: a) 50μm, b) 100μm, c) 150μm and 4) 200μm from the center of the hole.

This was also observed in simulations using Garfield++ [22] in which individual electrons were placed at four different distances from the center of the hole and above the THCOBRA foil (as indicated by the red dots in Fig. 9). A total of 1000 avalanches were simulated for each starting position. In Fig 9, the black dots represent the end point of each electron of the avalanche on a plane below the THCOBRA foil. The distribution of the final electron positions after the avalanche displays a dependence on the starting point and thus indicates a preservation of spatial information. While the full hole contributes for the case shown in Fig. 9a, almost no electrons are observed on the opposite side of the hole, when the arriving electron is outside of it, as shown in Fig. 9d. This is in agreement with the experimental observations presented above although a quantitative comparison might not be possible due to the widened incoming electron distribution in the experimental study.

This effect was also observed by using photons of 5.9keV from an $^{55}$Fe source and a high-sensitivity EM-CCD camera. The gas gain was increased and images were recorded as shown in Fig. 10. Recorded images display an asymmetry of the light emitted from these avalanches in the holes and anode strips, which indicates that the initial position of the incoming electrons for a single X-ray event is preserved up to some degree. However, it was not possible to disentangle the charge sharing contribution produced by the two thin GEMs located on top of the THCOBRA and the spread introduced by the THCOBRA foil itself. Further studies are needed to address this effect and quantify each contribution.

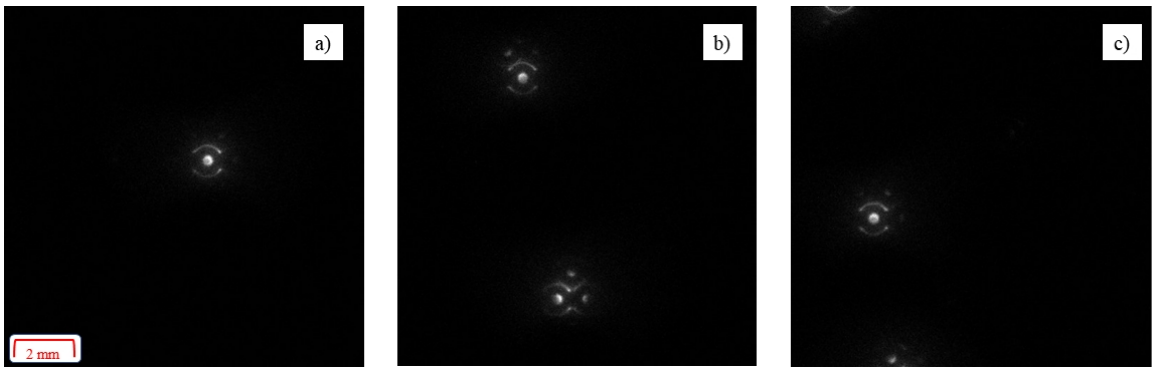

**Figure 10.** Images of individual low-energy X-ray photon interactions from an $^{55}$Fe source.

### 4.3 Source of instabilities

While the THCOBRA detector was operating mostly in stable conditions, occasional discharges were observed. In order to understand the origin of these discharges, multiple images of them were recorded and combined together to obtain the single image shown in Fig. 11.

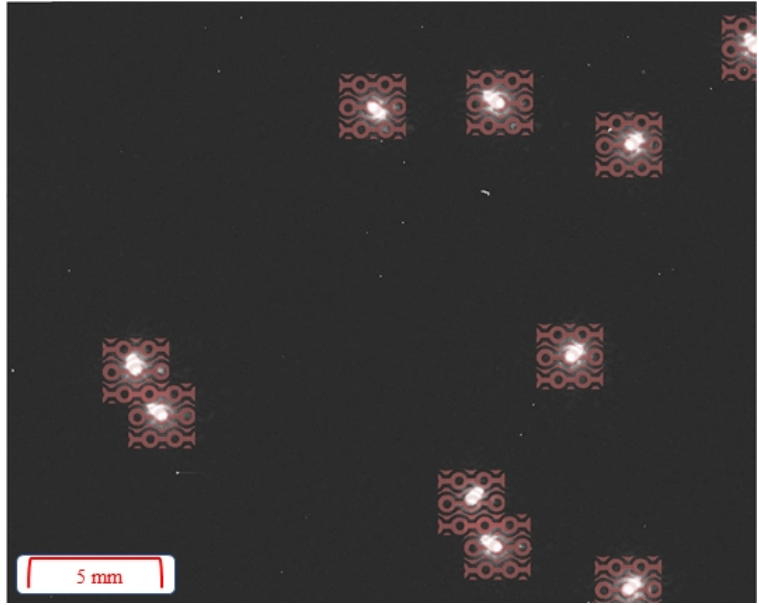

**Figure 11.** Image of discharges in different locations on the THCOBRA. The layout of the electrodes (red) was overlaid onto the image. Discharges are preferentially oriented towards the narrow regions and corners of the anode strips.

Discharges occurred at random locations across the THCOBRA foil, with a preferential orientation towards the narrow regions and corners of the anode strips, indicating the weak point of this electrode structure with an increased concentration of electric field lines.

## 5. Conclusions

In summary, this study allowed a deeper understanding of the THCOBRA operation and its limits along with a visualization of the sources of instabilities. The conclusions from this study are:

- The optical readout technique was successfully employed to read out a THCOBRA detector.

- The performance of this hybrid structure was visualized. The contributions of the holes were clearly separated from the contribution of the anode strips.

- The so-called "memory effect", i.e. a preservation of the spatial information from electrons reaching the THCOBRA, was observed. However, it was not possible to disentangle and quantify the contributions from charge sharing and this effect on the spatial resolution of the detector.

- The observation that scintillation light can be emitted from only certain regions within the holes (Fig. 8) as well as the computed position-dependent avalanche charge distributions (Fig. 9) show that the spatial position accuracy of THCOBRA structures can be better than the pitch of the holes, as reported in previous studies [21].

**Acknowledgments**

This work was partially supported by projects UID/CTM/50025/2019, PTDC/FIS-AQM/32536/2017 and CERN/FIS-INS/0025/2017 through COMPETE, Portugal, FEDER, Portugal and FCT programs, Portugal. Special thanks to Rui de Oliveira for the production and testing of the THCOBRA foils.

**References**

[1] F. D. Amaro et al 2010 JINST 5, P10002.

[2] F. Sauli, Nucl. Instrum. Meth. A386 (1997) 531.

[3] R. Chechik, et al., Nucl. Instrum. Meth. A535 (2004) 303.

[4] A. Breskin et al., Nucl. Instrum. Meth. A 598 (2009) 107.

[5] A. Oed, Nucl. Instrum. Meth. A263 (1988) 351.

[6] J.F.C.A. Veloso et al., Rev. Sci. Instrum. 71 (2000) 2371.

[7] Y. Giomataris et al., Nucl. Instrum. Meth. A376 (1996) 29.

[8] F.A.F. Fraga et al., Nucl. Instrum. Meth. A471 (2001) 125.

[9] M. Marafini et al (2015) JINST 10, P10034.

[10] M. Pfützner et al., Phys.Rev.C 90 (2014) 1, 014311.

[11] F. Brunbauer et al (2018) JINST 13, P11003.

[12] F. Brunbauer et al (2018) IEEE Trans. Nucl. Sci. 65(3), 903.

[13] T. Fujiwara et al., Nucl. Instrum. Meth. A850 (2017) 7.

[14] F.M. Brunbauer et al., Nucl. Instrum. Meth. A955 (2020) 163320.

[15] L. F. N. D. Carramate et al (2015) JINST 10, P01003.

[16] A. L. M. Silva et al (2015) Journal of Analytical Atomic Spectrometry 30(2), 343.

[17] L.F.M.D. Carramate et al (2017) JINST 12, T05003.

[18] F.A.F. Fraga et al., Nucl. Instrum. Meth. A504 (2003) 88.

[19] QImaging Retiga R6 CCD, viewed 20 July 2020, https://www.qimaging.com/retiga-r6

[20] Hamamatsu ImageEM EM-CCD camera, viewed 20 July 2020, https://www.hamamatsu.com/eu/en/product/type/C9100-23B/index.html

[21] A. L. M. Silva et al (2013) JINST 8, P05016.

[22] Rob Veenhof et al., Garfield++, viewed 20 July 2020, http://cern.ch/garfieldpp